\documentclass{INTERSPEECH2023}

\interspeechcameraready

\title{Improved Training for End-to-End Streaming Automatic Speech Recognition Model with Punctuation}
\name{Hanbyul Kim, Seunghyun Seo, Lukas Lee, Seolki Baek}
\address{
    NAVER Cloud, South Korea}
\email{
    \texttt{\{hanbyul.kim, real.seunghyun.seo, lukas.lee, seolki.baek\}@navercorp.com}
}

\usepackage{xcolor}
\usepackage{svg}

\usepackage{svg}
\usepackage{lipsum} 
\usepackage{algorithm}
\usepackage{algpseudocode}
\usepackage{float}
\usepackage{siunitx}

\begin{document}

\maketitle
 
\begin{abstract}
Punctuated text prediction is crucial for automatic speech recognition as it enhances readability and impacts downstream natural language processing tasks.
In streaming scenarios, the ability to predict punctuation in real-time is particularly desirable but presents a difficult technical challenge. 
In this work, we propose a method for predicting punctuated text from input speech using a chunk-based Transformer encoder trained with Connectionist Temporal Classification (CTC) loss.
The acoustic model trained with long sequences by concatenating the input and target sequences can learn punctuation marks attached to the end of sentences more effectively. 
Additionally, by combining CTC losses on the chunks and utterances, we achieved both the improved F1 score of punctuation prediction and Word Error Rate (WER).
\end{abstract}


\section{Introduction}
Advancements in deep neural networks have led to substantial improvements in Automatic Speech Recognition (ASR) systems. These systems are extensively utilized in various applications, such as voice assistants and dictation software. 
In recent years, Transformer models \cite{tfm}, self-attention-based neural networks, have exhibited remarkable performance to extract high-level speech representations in ASR.
Self-attention mechanism utilized in these models exploits contextual information for acoustic encoding by calculating attention weights for each pair of input features, allowing the modeling of temporal dependencies within a sequence.
However, due to its requirement of having the entire input sequence available before processing, the limitation prevents it from being a suitable encoder for streaming ASR applications, which require the real-time recognition of spoken words shortly after they are uttered.
This study focuses on Connectionist Temporal Classification (CTC) based streaming models, despite the existence of alternative methods such as RNN-T (Recurrent Neural Network Transducer) \cite{he2019streaming, zhang2020transformer} or encoder-decoder \cite{watanabe2017hybrid, moritz2020streaming} based models.

Several approaches can be employed to enable the utilization of transformer-based models for streaming scenarios.
Time-restricted self-attention layer \cite{trtfm} was proposed to limit the number of look-ahead frames, alleviating the requirement of complete sequences.
However, as the receptive field on the future frames scales linearly with the number of transformer layers, it results in significant latency for a deep architecture.
An alternative approach is a chunk-based method, which divides the input signal into shorter segments and feeds them to the model sequentially. 
This method reduces the computational cost and latency compared to the full-sequence model while maintaining comparable performance. 
Recent studies \cite{miao2020transformer, an2020cat, wu2020streaming} have shown that the chunk-based method generally outperforms time-restricted self-attention regarding both accuracy and latency reduction, due to its ability to leverage contextual information within each segment better. 




Many studies have attempted to enhance the accuracy of speech recognition systems, but relatively few have given sufficient attention to punctuation accuracy. 
Punctuation marks play a crucial role in enhancing the readability and comprehension of the transcriptions.
Moreover, the accuracy of punctuation prediction can have significant implications for downstream natural language processing tasks. 
Previous studies \cite{tilk2016bidirectional, fang2019using, reviewer7} have assumed a cascaded system of two separate models for ASR and punctuation prediction.
However, incorporating a punctuation model, such as BERT \cite{devlin-etal-2019-bert} into ASR systems poses several challenges. 
Firstly, pre-trained BERT can perform poorly because ASR output is fed to BERT, where ASR output has acoustic noises that BERT has never seen in training time.
In addition, cascaded systems might introduce additional latency in streaming scenarios. 
Punctuation models generally cannot make predictions until the entire input sequence is received. 
Moreover, the incorporation of large neural networks in these systems may result in additional computational burdens.
Due to these reasons, the end-to-end approach for formatted speech recognition is gaining attention from researchers \cite{mimura2021end, s2pt}, but studies for streaming models still lack.

The task of end-to-end punctuation prediction in streaming ASR presents a unique challenge compared to non-streaming ASR. 
Most non-streaming ASR models are trained to accept full utterances with speech segments provided by additional modules, such as a Voice Activity Detection (VAD) model. 
Some key punctuation marks, such as period and question, are typically located at the end of sentences. 
Therefore, proper segmentation of the speech input makes the prediction of these key punctuation marks relatively straightforward, as reported in the previous study \cite{s2pt}.
Nonetheless, recent studies about end-to-end VAD approach \cite{vad} or punctuation prediction with acoustic features \cite{christensen2001punctuation, sunkara2020multimodal} 
proposed joint training approaches using both acoustic and semantic information.

\begin{figure*}[ht]
\centering
\includegraphics[height=6.0cm]{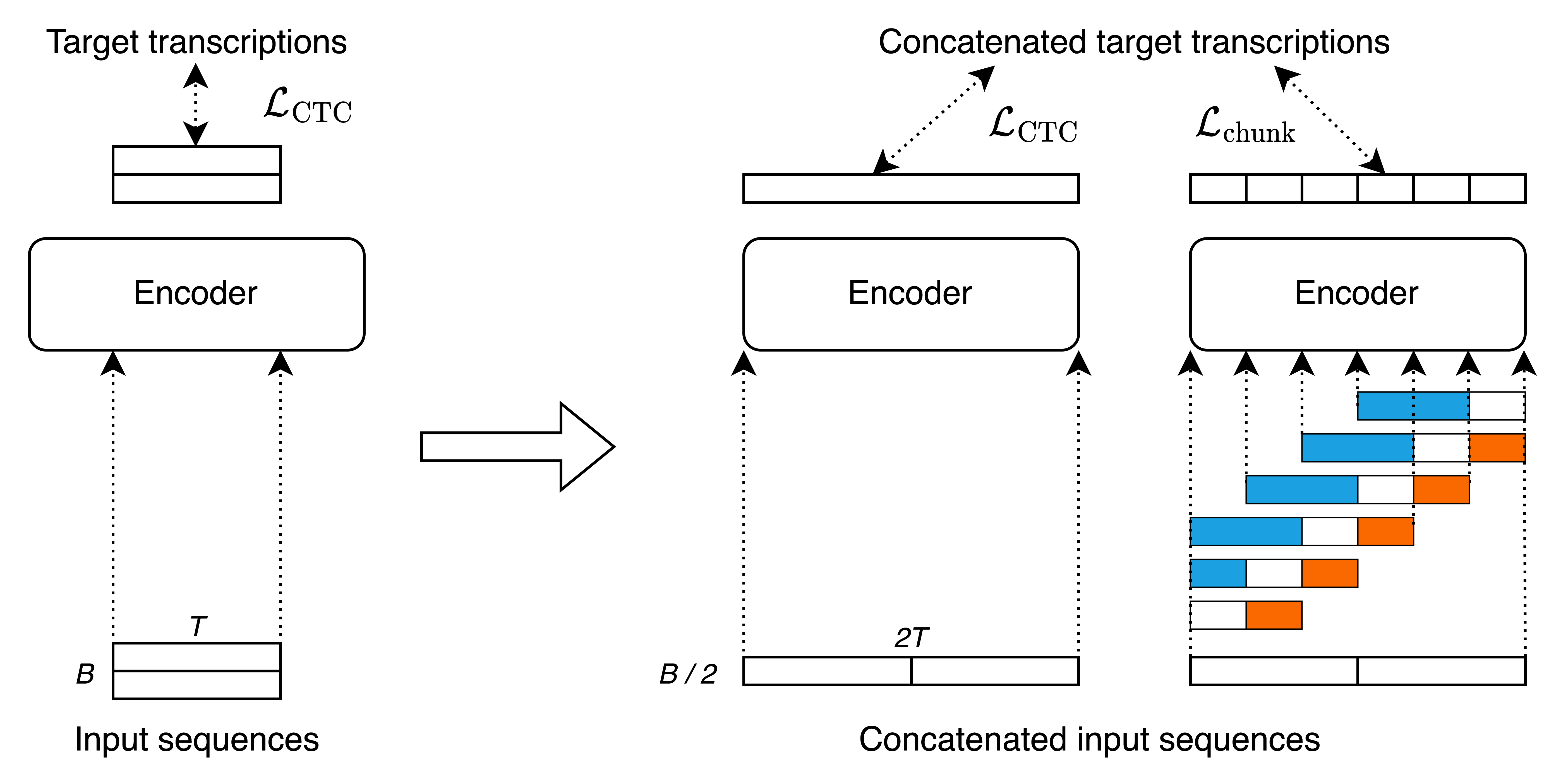}
\caption{Overview of the proposed method for end-to-end streaming ASR with punctuation prediction. The approach  concatenates input speech sequences and target text sequences for a CTC model to learn punctuation marks between the target labels. This doubles the number of frames while reducing the batch size by half. An additional CTC loss term is introduced, which is computed on the concatenated emissions of separated chunks. The blue and orange boxes in the figure represent past and future frames respectively, providing contextual information for a input sequence. The ASR model is trained by jointly optimizing CTC losses on full sequences and chunks.}
\label{fig:overview} 
\end{figure*}

In this paper, we propose a novel training method to enhance the accuracy of end-to-end streaming ASR models with punctuation prediction. 
Our approach is specifically designed for streaming ASR, where we utilize the transformer model trained with CTC loss. 
However, as we described above, predicting punctuation directly with a streaming ASR model poses challenges that are not present in non-streaming ASR. 
To address the issues, we present a technique that involves feeding concatenated speech mini-batches to a chunk-based transformer, which allows the model to learn the positions of punctuation marks in the middle of the sequence, rather than at the end.
To ensure stable speech recognition performance, we introduce an auxiliary loss that focuses on the chunk-hopping method.
Our experiments demonstrate that this training method improves punctuation prediction in an end-to-end manner, while also enhancing speech recognition accuracy in streaming scenarios.

\section{Related Work}

\subsection{Chunk-based streaming ASR}
Chunk-based streaming ASR is an approach that aims to reduce latency in ASR systems.
This method involves splitting the speech into chunks and processing each chunk using the ASR model, without waiting for the entire utterance to be completed \cite{li2015constructing, zeyer2016towards}. 
This allows the system to begin producing output in real time, providing users with faster and more efficient speech recognition capabilities. 
The chunk-based approach has been further developed to include context-sensitive chunks \cite{chen2016training}, where each chunk is encoded with additional left and right frames to provide the speech context. 
This approach has been explored in many studies \cite{dong2019self, miao2020transformer, an2020cat}, and found to be particularly effective when used with transformer-based ASR models.

\subsection{Joint training in streaming ASR}
ASR models with limited context can learn from a broader range of contextual information and improve their overall recognition performance by incorporating auxiliary loss functions in addition to the primary objective.
This approach may be particularly effective for streaming ASR, as it only increases computational burden during the training phase and enables the model to maintain low latency during recognition.

For encoder-decoder models, the online CTC/attention mechanism \cite{9003920, miao2020online, zhang2020streaming} is a technique that combines both the attention loss from the decoder and the CTC loss from the encoder.
Additionally, a ``soft forgetting'' approach \cite{audhkhasi2019forget, an2020cat} aims to minimize the mean squared error between the full utterances and the chunk-based encoder output to improve the model's performance on streaming ASR tasks.
The integration of streaming and non-streaming ASR systems by weight sharing and joint training a streaming model with a full-context model \cite{dualmodeasr, cuside} has emerged as a recent approach for improving the accuracy and low-latency performance of streaming ASR.

\section{Proposed Method}

In this paper, we propose an ASR encoder that can directly produce punctuated text from an acoustic input sequence.
We utilized a Transformer encoder trained with CTC loss, which is suitable for streaming ASR.
The CTC objective, first proposed by Graves, et al. \cite{graves2006connectionist} is a loss function that is commonly used in speech recognition tasks. 
It is designed to solve the problem of aligning input sequences and output sequences with different lengths in speech recognition due to the variability in speech duration and pronunciation. 

Given a sequence of speech inputs $\mathbf{X}=(\mathbf{x_1},\cdots, \mathbf{x_N})$ and transcriptions $\mathbf{Y}=(\mathbf{y_1},\cdots, \mathbf{y_N})$, the objective computes the likelihood of all possible alignments using representation vectors from ASR model and target sequence vectors. 

\begin{equation*}
P_{\text{CTC}}(\mathbf{y_i} \vert \mathbf{x_i}) = \sum_{\pi \in \beta^{-1}(\mathbf{y_i})} P( \pi \vert \mathbf{x_i} )
\label{eq:ctc_prob}
\end{equation*}

\vspace{-1mm}
\begin{equation*}
\mathcal{L}_{\text{CTC}} ( \mathbf{X}, \mathbf{Y} ) = - \sum_{i=1}^{N} \log P_{\text{CTC}}(\mathbf{y_i} | \mathbf{x_i})
\label{eq:ctc_loss}
\end{equation*}
where $\beta^{-1}(y)$ is the set of all possible alignments between input sequences and target sequences.

The proposed method involves modifying the input sequence to enable the model to effectively learn to predict punctuation marks.
First, given a current mini-batch, a new batch is created by concatenating two neighboring speech inputs and labels each from the current batch.
In most publicly available speech recognition datasets, punctuation marks such as periods and question marks are conventionally placed at the end of sentences. 
This may lead to inconsistencies, as punctuation marks can appear at any position within a chunk of speech data.
It should be noted that the adjacent speech inputs being concatenated would be semantically unrelated because we use randomly shuffled mini-batches.
\vspace{-0.1mm}
\begin{equation*}
\mathbf{\tilde{X}} = (\mathbf{\tilde{x_1}}, \cdots, \mathbf{\tilde{x_{N/2}}}) = (cat[\mathbf{x_1},\mathbf{x_2}], \cdots, cat[ \mathbf{x_{N-1}},\mathbf{x_{N}}] )
\label{eq:concatX}
\end{equation*}

\vspace{-6mm}
\begin{equation*}
\mathbf{\tilde{Y}} = (\mathbf{\tilde{y_1}}, \cdots, \mathbf{\tilde{y_{N/2}}}) = (cat[\mathbf{y_1},\mathbf{y_2}], \cdots, cat[ \mathbf{y_{N-1}},\mathbf{y_{N}}] )
\label{eq:concatY}
\end{equation*}
where $cat[\cdot]$ is a function that concatenates the two inputs in the time direction. 
From this concatenated batch, the CTC posterior probability can be computed.


The new batch tensor is split into multiple chunks in the time dimension, which is then padded with the left and right chunks of \si{2s} and \si{1s} respectively.
This is called the context-sensitive chunk technique to provide contextual information to chunks \cite{audhkhasi2019forget, an2020cat, cuside}. 
The ASR model takes each padded chunk as input separately and outputs representation vectors of each.
Among CTC emissions from each chunk, past and future frames are discarded and the remaining emissions are merged to obtain the final posterior which has the same length as the concatenated sequences.

\vspace{-3mm}
\begin{equation*}
P_{\text{chunk}}(\mathbf{\tilde{y_i}} \vert \mathbf{\tilde{x_i}}) = \sum_{\pi \in \beta^{-1}(\mathbf{y_i})}  P( \pi \vert \mathbf{\tilde{x_{i1}}}, \cdots, \mathbf{\tilde{x_{ic}}} )
\label{eq:chunk_prob}
\end{equation*}
where $\mathbf{c}$ is the number of chunks and $\mathbf{\tilde{x_{i1}}}, \cdots, \mathbf{\tilde{x_{ic}}}$ are chunks of concatenated utterance.
Note that, unlike local self-attention, this method does not suffer from the disadvantage of a larger receptive field as the layer gets deeper.

From each posterior probability, $P_{\text{CTC}}$ and $P_{\text{chunk}}$, corresponding CTC losses $\mathcal{L}_{\text{CTC}}$ and $ \mathcal{L}_{\text{chunk}}$ can be computed.
Finally, the total loss is obtained by interpolating two losses.


\vspace{-3.5mm}
\begin{equation*}
\mathcal{L}_{\text{total}} ( \mathbf{X}, \mathbf{Y} ) = (1 - \lambda) \mathcal{L}_{\text{CTC}} ( \mathbf{\tilde{X}}, \mathbf{\tilde{Y}} ) + \lambda \mathcal{L}_{\text{chunk}} ( \mathbf{\tilde{X}}, \mathbf{\tilde{Y}} )
\label{eq:total_loss}
\end{equation*}

The detailed training procedure is outlined in Algorithm \ref{alg:alg1}.

\begin{algorithm}[h]
\caption{Pseudo-code of the proposed method}
\label{alg:alg1}
\begin{algorithmic}
\Procedure{TrainStep}{X, Y}
\State $X_{concat} \gets \Call{Concat}{X}$ \Comment{Concatenate input speech}
\State $Y_{concat} \gets \Call{Concat}{Y}$ \Comment{Concatenate target labels} \\
\State $P_{CTC} \gets \Call{Encoder}{X_{concat}}$
\State $L_{CTC} \gets \Call{CTC}{P_{CTC}, Y_{concat}}$ \\
\State $[X_{split}]^{N}_{1} \gets \Call{Split}{X_{concat}}$ \Comment{Split input speech into non-overlapping chunks}
\State $[X_{pad}]^{N}_{1} \gets \Call{PadContext}{[X_{split}]^{N}_{1}}$ \Comment{Pad the past and future frames for context}
\State $[P_{pad}]^{N}_{1} \gets \Call{Encoder}{[X_{pad}]^{N}_{1}}$
\State $[P_{chunk}]^{N}_{1} \gets \Call{RemoveContext}{[P_{pad}]^{N}_{1}}$ \Comment{Remove the padded frames}
\State $P_{merge} \gets \Call{Merge}{[P_{chunk}]^{N}_{1}}$
\State $L_{chunk} \gets \Call{CTC}{P_{merge}, Y_{concat}}$
\State $L_{total} \gets (1-\lambda) * L_{CTC} + \lambda * L_{chunk}$
\State $\Call{Backward}{L_{total}}$
\EndProcedure
\end{algorithmic}
\end{algorithm}

\begin{table*}[ht]
\centering
\begin{tabular}{@{}lrrrrrrrrrrrrr@{}}
\toprule
\multicolumn{1}{c}{}      & \multicolumn{1}{c}{}         & \multicolumn{4}{c}{Precision score (\%)}                                                         & \multicolumn{4}{c}{Recall score (\%)}                                                            & \multicolumn{4}{c}{F1 score (\%)}                                                                \\
\multicolumn{1}{c}{Model} & \multicolumn{1}{c}{WER (\%)} & \multicolumn{1}{c}{,} & \multicolumn{1}{c}{.} & \multicolumn{1}{c}{?} & \multicolumn{1}{c}{avg.} & \multicolumn{1}{c}{,} & \multicolumn{1}{c}{.} & \multicolumn{1}{c}{?} & \multicolumn{1}{c}{avg.} & \multicolumn{1}{c}{,} & \multicolumn{1}{c}{.} & \multicolumn{1}{c}{?} & \multicolumn{1}{c}{avg.} \\ \midrule
Baseline                  & 18.2                         & 40.4                  & 73.5                  & 52.7                  & 55.5                     & 66.6                  & 23.6                  & 25.1                  & 38.4                     & 50.3                  & 35.8                  & 34.0                  & 40.0                     \\
\hspace{0.2cm}+ interCTC \cite{s2pt}               & 18.5                         & 36.3                  & 73.2                  & 50.6                  & 53.4                     & 69.2                  & 21.2                  & 19.1                  & 36.5                     & 47.6                  & 32.8                  & 27.8                  & 36.1                     \\
Proposed                  & 15.8                         & 39.0                  & 76.0                  & 53.5                  & 56.2                     & 72.8                  & 33.6                  & 29.4                  & 45.3                     & 50.8                  & 46.6                  & 37.9                  & 45.1                     \\ \bottomrule
\end{tabular}

\caption{WER and punctuation prediction accuracy on long-form utterances in MuST-C testset}
\label{tab:t1}
\vspace{-3.5mm}
\end{table*}

\section{Experiments}

\subsection{Dataset}
MuST-C dataset \cite{mustc} was utilized for both training and evaluating the efficacy of our approach.
MuST-C is a multilingual speech translation corpus that includes at least 385 hours of audio recordings from English TED Talks for each of the eight target languages.
We obtained data from the English-German speech translation dataset by extracting both English speech and its corresponding script, including punctuation.

The following processing steps were performed on the data labels. 
All labels were converted to lowercase to maintain consistency and reduce complexity. 
Punctuation marks, with the exception of apostrophes('), commas(,), periods(.), and question marks(?), were removed to simplify the text while preserving contextual information.
Environmental labels enclosed in brackets (e.g. (Applause), (Laughter)) and speaker names followed by a colon (e.g. MJ:) were removed to prevent the ASR model from predicting them as output.
The original train and dev sets were used in the same configuration for training and validation, respectively.

\subsection{Training Setup}
The training process for the proposed ASR model was conducted using the fairseq \cite{fairseq} framework. 
The ASR model architecture consisted of a Transformer with 12 encoder layers, 256 hidden dimensions, and 4 attention heads.
We added subsampling layers to downsample the input features. 
The layers consisted of two convolutional layers with the first layer having an input channel size of 80 and an output channel size of 1024, and the second layer having an input channel size of 512 and an output channel size of 512. 
Both layers had a kernel size of 5 and a stride of 2.
For streaming ASR, it is natural to use relative position encoding instead of absolute position encoding.
Hence, we replaced the absolute position encoding layer with a convolutional positional encoding layer.
The convolutional position encoding layer was implemented using a conv1d layer with a kernel size of 128, 256 output channels, and 16 groups. This implementation is consistent with the approach used in \cite{w2v2}.
The total number of parameters in the model is approximately 18 million.

During training, the CTC loss with auxiliary chunk loss was used as the loss function, with interpolating factor $\lambda$ set to 0.5.
The input feature consisted of 80-dimensional log-mel filterbanks. 
The label data was encoded with 2000 wordpieces, which were trained using the label data of the training set with SentencePiece \cite{kudo2018sentencepiece}. 
To optimize training speed and reduce memory consumption, the model was trained with half-precision floating-point (FP16). 
Adam \cite{adam} was chosen as the optimization algorithm for this study.
The learning rate was set to 0.001, and the learning rate scheduler was the inverse square root.
The warm-up steps were set to 10000, and the model was trained for 190 epochs, with approximately 100,000 training steps. 
The maximum token size was set to 40000 tokens, corresponding to a batch size of approximately 400. 
The training was conducted using NVIDIA A100 8 GPUs, each equipped with 80 GB of VRAM.

\subsection{Evaluation}
In this study, we used ``tst-COMMON'' subset of MuST-C dataset to evaluate our approach.
The dataset is based on long English TED-talk recordings that consisted of audio transcriptions with corresponding timestamps. 
This enabled us to evaluate the proposed method on both long-form utterances using the original audio files, as well as short-form utterances by splitting the speech data according to the timestamps. 
The average duration of complete utterances was found to be 11 minutes, while the average duration of the individual segments obtained after splitting them was 5.8 seconds.

We consisted of a chunk size of \si{1s} with \si{2s} of padding for past and \si{1s} for future frames, consistent with the model training, during the chunk-encoding process. 
After computing the emission from the transformer encoder, we performed CTC beam search with a beam size of 5, without using any additional language model. 
For evaluating ASR accuracy, we measured the word error rate (WER), excluding any punctuation text.
Additionally, we measured precision, recall and F1 score for each punctuation mark to evaluate the punctuation accuracy. 
Due to ASR errors, we first aligned the ASR prediction and ground truth and then calculated the F1 score.

\subsection{Experimental Results}
Table \ref{tab:t1} shows WER results for streaming speech recognition accuracy. 
The proposed method outperforms the baseline method by 13\% relatively (18.2 vs. 15.8).
In addition, we compared the punctuation recognition accuracy, and the proposed method demonstrated 12.5\% higher average F1 scores across all punctuation marks.
Further analysis of precision and recall reveals that the proposed method significantly improves the recall score. 
The recall of the period prediction shows a significant improvement with a nearly 50\% increase (23.6 vs. 33.6).
This suggests that our method better predicts the end of a sentence.
We compared our proposed approach with the existing method \cite{s2pt} suggested previously.
To reproduce the approach, we introduced an intermediate CTC loss with unpunctuated labels, which was jointly trained with the CTC loss. 
However, the proposed approach does not show any improvement in either WER or punctuation scores in streaming ASR.

Table \ref{tab:short_eval} presents the evaluation results of previous studies and the proposed method on segmented speech with shorter lengths. Overall, the WER slightly degrades across all approaches. Despite the significant drop in F1 score on period mark compared to long-form decoding, the proposed approach demonstrates superior performance with the highest average punctuation prediction accuracy among all approaches.

\subsection{Analysis} 
In this section, we analyzed the factors that contributed to the effectiveness of our proposed approach.
To determine the impact of specific factors, we re-trained our model with various modifications to our original configuration. 
Table \ref{tab:t2} illustrates the effects of the modifications.

First, we removed the chunk CTC loss ($\mathcal{L}_{\text{chunk}}$) and trained the model again with concatenated input sequences only.
We found that WER increased significantly, indicating that proper regularization with the chunk CTC loss is crucial to prevent accuracy degradation.
Additionally, we observed that our proposed method significantly boosted the accuracy of period marks, exceeding the improvement observed with the proposed approach.
However, this has a negative effect on commas and question marks, leading to a lower average F1 score compared to our method. 
From these observations, we conclude that chunk CTC loss not only regulates the CTC model to prevent WER degradation but also prevents period marks over-fitting while maintaining accuracy for other punctuation marks.

\begin{table}[ht!]
\centering
\begin{tabular}{@{}lrrrrr@{}}
\toprule
\multicolumn{1}{c}{}         & \multicolumn{1}{c}{}         & \multicolumn{4}{c}{Punctuation F1 score (\%)}                                                    \\
\multicolumn{1}{c}{Model}    & \multicolumn{1}{c}{WER (\%)} & \multicolumn{1}{c}{,} & \multicolumn{1}{c}{.} & \multicolumn{1}{c}{?} & \multicolumn{1}{c}{avg.} \\ \midrule
Baseline                     & 18.5                         & 47.7                  & 16.0                  & 39.4                  & 34.4                     \\
\hspace{0.2cm}+ interCTC \cite{s2pt} & 18.9                         & 44.6                  & 12.5                  & 29.1                  & 28.7                     \\
Proposed                     & 16.3                         & 52.5                  & 27.3                  & 44.0                  & 41.2                     \\ \bottomrule
\end{tabular}
\caption{Performance evaluation of WER and punctuation prediction accuracy on speech segments in MuST-C testset}
\label{tab:short_eval}
\end{table}
\vspace{-8mm}
\begin{table}[ht!]
\centering
\begin{tabular}{@{}lrrrrr@{}}
\toprule
\multicolumn{1}{c}{}      & \multicolumn{1}{c}{}         & \multicolumn{4}{c}{Punctuation F1 score (\%)}                                                    \\
\multicolumn{1}{c}{Model} & \multicolumn{1}{c}{WER (\%)} & \multicolumn{1}{c}{,} & \multicolumn{1}{c}{.} & \multicolumn{1}{c}{?} & \multicolumn{1}{c}{avg.} \\ \midrule
Proposed                  & 15.8                         & 50.8                  & 46.6                  & 37.9                  & 45.1                     \\
\hspace{0.2cm}- $\mathcal{L}_{\text{chunk}}$                & 21.3                         & 45.0                  & 52.9                  & 25.8                  & 41.2                     \\ 
\hspace{0.2cm}- concat.                  & 16.0                         & 50.1                  & 41.0                  & 34.0                  & 41.7                     \\
\bottomrule
\end{tabular}
\caption{Ablation study of the effect of auxiliary loss term on chunk-based output and input sequence concatenation}
\label{tab:t2}
\end{table}
\vspace{-3mm}

We also conducted an experiment to assess the impact of excluding the input sequence concatenation in our training approach.
The results show that the overall F1 scores for punctuation marks drop, which suggests that concatenating input sequences plays a crucial role in improving the punctuation recognition performance of our approach.
Moreover, an interesting observation was that removing the concatenation also degraded the WER slightly. 
This could be due to the fact that the CTC loss computed on concatenated sequences also regularizes the joint training approach, indicating the importance of the regularization effect.
These results highlight the effectiveness of our training method which improves punctuation recognition accuracy while maintaining the streaming ASR performance.

\section{Conclusion}
This work proposed a novel training method to improve both speech recognition and punctuation prediction performances in the streaming ASR task.
The experimental results showed that the concatenation of input sequences played a significant role in improving the punctuation recognition performance of the proposed method.
The chunk-based loss was crucial not only for preventing WER degradation by input concatenation but also for preventing period mark over-fitting while maintaining accuracy for other punctuation marks.
While our approach does not introduce additional computation burden during inference, the training time and space complexity increased due to the chunk-wise forward steps. 
Further experimentation is required to optimize the training computation for our approach.
We also would like to investigate the applicability of the proposed approach to other ASR models and datasets.

\newpage

\bibliographystyle{IEEEtran}
\bibliography{mybib}

\end{document}